\documentclass[conference]{IEEEtran}

\usepackage[utf8]{inputenc}
\usepackage{amsmath, amssymb, amsfonts}
\usepackage{graphicx}
\usepackage{booktabs}
\usepackage{multirow}
\usepackage{array}
\usepackage{algorithm}
\usepackage{algorithmic}
\usepackage[hidelinks]{hyperref}
\usepackage{cite}
\usepackage{caption}
\usepackage{subcaption}

\begin{document}

\title{Illuminating the Black Box: Real-Time Monitoring of Backdoor Unlearning in CNNs via Explainable AI}

\author{\IEEEauthorblockN{Hoang Tien Dat}
\IEEEauthorblockA{Department of Computer Science\\
Phenikaa University\\
Hanoi, Vietnam\\
Email: itentad.work@gmail.com}}

\maketitle

\begin{abstract}
Backdoor attacks pose severe security threats to deep neural networks by embedding malicious triggers that force misclassification. While machine unlearning techniques can remove backdoor behaviors, current methods lack transparency and real-time interpretability. This paper introduces a novel framework that integrates Gradient-weighted Class Activation Mapping (Grad-CAM) into the unlearning process to provide real-time monitoring and explainability. We propose the \textit{Trigger Attention Ratio} (TAR) metric to quantitatively measure the model's attention shift from trigger patterns to legitimate object features. Our balanced unlearning strategy combines gradient ascent on backdoor samples, Elastic Weight Consolidation (EWC) for catastrophic forgetting prevention, and a recovery phase for clean accuracy restoration. Experiments on CIFAR-10 with BadNets attacks demonstrate that our approach reduces Attack Success Rate (ASR) from 96.51\% to 5.52\% while retaining 99.48\% of clean accuracy (82.06\%), achieving a 94.28\% ASR reduction. The integration of explainable AI enables transparent, observable, and verifiable backdoor removal.
\end{abstract}


\section{Introduction}

Deep neural networks (DNNs) have achieved remarkable success across computer vision, natural language processing, and various safety-critical applications. However, their vulnerability to backdoor attacks has emerged as a critical security concern. In a backdoor attack, an adversary poisons a small fraction of the training data by embedding a specific trigger pattern and associating it with a target label. The compromised model performs normally on clean inputs but consistently misclassifies any input containing the trigger to the attacker's desired class~\cite{gu2017badnets}.

Traditional backdoor defense mechanisms, including neural cleanse~\cite{wang2019neural}, fine-pruning~\cite{liu2018fine}, and neural attention distillation~\cite{li2021neural}, often require retraining or significant architectural modifications. More critically, these approaches operate as ``black boxes,'' providing no visibility into the unlearning process or verification that backdoor behaviors are actually being removed rather than simply suppressed.

\subsection{Motivation}

The lack of transparency in backdoor removal creates several fundamental problems:

\begin{enumerate}
    \item \textbf{Verification Challenge}: Security practitioners cannot verify whether the backdoor has been genuinely eliminated or merely hidden deeper in the model's latent space.
    
    \item \textbf{Trust Deficit}: Without observable evidence of backdoor removal, deploying supposedly ``cleaned'' models in critical systems remains risky.
    
    \item \textbf{Debugging Difficulty}: When unlearning fails, practitioners lack diagnostic tools to understand why the process was unsuccessful.
\end{enumerate}

Recent advances in Explainable AI (XAI), particularly Grad-CAM~\cite{selvaraju2017grad}, offer a promising solution. However, existing research has only applied XAI techniques for \textit{post-hoc analysis} after unlearning completes, rather than integrating them directly into the unlearning loop for real-time monitoring.

\subsection{Research Gap}

Current machine unlearning approaches for backdoor removal suffer from three critical limitations:

\textbf{Limited Transparency}: Existing methods like gradient ascent and knowledge distillation modify model weights without providing interpretable feedback on what is being unlearned~\cite{bourtoule2021machine}.

\textbf{No Real-Time Monitoring}: Practitioners must wait until training completes to evaluate whether backdoor removal succeeded, wasting computational resources on failed attempts.

\textbf{Lack of Quantitative Metrics}: Beyond accuracy-based metrics (ASR, clean accuracy), there are no specialized measurements quantifying the model's attention shift away from trigger patterns.

\subsection{Contributions}

This paper makes three primary contributions:

\textbf{(1) Real-Time XAI Monitoring Framework}: We integrate Grad-CAM directly into the unlearning training loop, generating activation heatmaps at regular intervals to visualize attention patterns as backdoor behaviors are removed.

\textbf{(2) Trigger Attention Ratio (TAR) Metric}: We introduce TAR, a novel quantitative metric that measures the ratio of model attention focused on trigger regions versus object regions:
\begin{equation}
\text{TAR} = \frac{\sum_{(i,j) \in \mathcal{R}_{\text{trig}}} M_{i,j}}{\sum_{(i,j) \in \mathcal{R}_{\text{obj}}} M_{i,j} + \epsilon}
\label{eq:tar}
\end{equation}
where $M$ is the Grad-CAM heatmap, $\mathcal{R}_{\text{trig}}$ is the trigger region, $\mathcal{R}_{\text{obj}}$ is the object region, and $\epsilon$ prevents division by zero.

\textbf{(3) Balanced Unlearning Strategy}: We propose a three-component loss function that balances backdoor removal (gradient ascent), clean accuracy preservation (supervised learning), and weight stability (EWC regularization), plus a recovery phase that restores clean accuracy while maintaining low ASR.

Our experimental results on CIFAR-10 demonstrate a 94.28\% reduction in ASR (96.51\% $\rightarrow$ 5.52\%) while retaining 99.48\% of the original clean accuracy (82.06\%).

\section{Related Work}

\subsection{Backdoor Attacks and Defenses}

Backdoor attacks on neural networks were first systematically studied by Gu et al.~\cite{gu2017badnets}, who demonstrated that data poisoning during training can embed persistent malicious behaviors.

\textbf{Neural Cleanse}~\cite{wang2019neural} formulates backdoor detection as an optimization problem, reverse-engineering potential triggers. While effective at detection, it requires extensive computation and cannot directly remove backdoors.

\textbf{Fine-Pruning}~\cite{liu2018fine} identifies and removes neurons most activated by poisoned samples. However, this approach may damage model accuracy and cannot guarantee complete backdoor removal if the backdoor is distributed across multiple neurons.

\textbf{Neural Attention Distillation (NAD)}~\cite{li2021neural} uses knowledge distillation to transfer attention from backdoored to clean models. NAD achieves strong results (clean accuracy $\sim$90\%, ASR $\sim$7.2\%) on CIFAR-10 with ResNet-18, representing current state-of-the-art for backdoor unlearning. However, NAD requires a clean teacher model and provides no interpretable feedback during the unlearning process.

\subsection{Machine Unlearning}

Machine unlearning~\cite{bourtoule2021machine} aims to selectively remove specific training samples' influence from a trained model without full retraining.

\textbf{Gradient Ascent Methods} maximize loss on backdoor samples to reverse their learned influence. However, naive gradient ascent often causes catastrophic forgetting.

\textbf{Elastic Weight Consolidation (EWC)}~\cite{kirkpatrick2017overcoming} addresses catastrophic forgetting by adding a regularization term:
\begin{equation}
\mathcal{L}_{\text{EWC}} = \mathcal{L}_{\text{task}} + \frac{\lambda}{2} \sum_{i} F_i (\theta_i - \theta_i^*)^2
\end{equation}

\subsection{Explainable AI in Security}

Grad-CAM~\cite{selvaraju2017grad} computes class-discriminative localization maps:
\begin{equation}
L_c^{\text{Grad-CAM}} = \text{ReLU}\left(\sum_k \alpha_k^c A^k\right)
\end{equation}

Recent work has used Grad-CAM to analyze backdoor triggers~\cite{wang2019neural}, but only for \textit{post-hoc analysis}. Our work bridges this gap by integrating Grad-CAM into the unlearning training loop.

\section{Methodology}

\subsection{Threat Model}

We consider the BadNets attack scenario~\cite{gu2017badnets} on CIFAR-10:

\textbf{Dataset}: 60,000 color images (32$\times$32 pixels) across 10 classes, split into 50,000 training and 10,000 test images.

\textbf{Backdoor Trigger}: A 4$\times$4 white square patch at the bottom-right corner:
\begin{equation}
\mathbf{x}_{\text{trigger}}[i,j,:] = 1.0 \quad \forall (i,j) \in [28,31] \times [28,31]
\end{equation}

\textbf{Poison Rate}: 3\% of training data (1,500 images).

\textbf{Target Class}: Class 0 (airplane).

\subsection{Model Architecture}

We employ a custom CNN architecture:

\textbf{Block 1}: Two Conv2D layers (64 filters, 3$\times$3, ReLU) + MaxPooling (2$\times$2).

\textbf{Block 2}: Two Conv2D layers (128 filters, 3$\times$3, ReLU) + MaxPooling (2$\times$2).

\textbf{Block 3}: One Conv2D layer (256 filters, 3$\times$3, ReLU, named ``last\_conv'') + Global Average Pooling.

\textbf{Classifier}: Dense(256, ReLU) + Dense(10, softmax).

The model contains approximately 623,690 trainable parameters, achieving 82.49\% clean accuracy and 96.51\% ASR when trained on poisoned data.

\subsection{Proposed Framework}

\subsubsection{Phase 1: Unlearning with Combined Loss}

The core of our approach is a combined loss function:

\begin{equation}
\mathcal{L}_{\text{total}} = -\alpha_t \mathcal{L}_{\text{poison}} + \lambda \mathcal{L}_{\text{clean}} + \beta \|\boldsymbol{\theta} - \boldsymbol{\theta}_0\|^2
\label{eq:combined_loss}
\end{equation}

where:
\begin{itemize}
    \item $\mathcal{L}_{\text{poison}}$ is cross-entropy loss on poisoned samples
    \item $\mathcal{L}_{\text{clean}}$ is cross-entropy loss on clean samples
    \item $\alpha_t$ is the dynamic gradient ascent weight (0.05 $\rightarrow$ 0.20)
    \item $\lambda = 1.0$ controls clean loss importance
    \item $\beta = 1 \times 10^{-4}$ is the EWC regularization coefficient
\end{itemize}

\textbf{Alpha Scheduling}: $\alpha_t$ increases linearly over 15 epochs:
\begin{equation}
\alpha_t = \alpha_{\text{start}} + \frac{t-1}{T-1}(\alpha_{\text{end}} - \alpha_{\text{start}})
\end{equation}

\subsubsection{Phase 2: Recovery Phase}

After aggressive unlearning, we fine-tune on clean data:
\begin{equation}
\mathcal{L}_{\text{recovery}} = \mathcal{L}_{\text{clean}} + \beta \|\boldsymbol{\theta} - \boldsymbol{\theta}_{\text{unl}}\|^2
\end{equation}

\subsubsection{Phase 3: Real-Time Monitoring}

\textbf{Periodic Visualization}: Every epoch, we sample 5 poisoned test images and generate Grad-CAM heatmaps.

\textbf{TAR Computation}: For each heatmap $M \in \mathbb{R}^{H \times W}$:
\begin{equation}
\text{TAR} = \frac{\sum_{(i,j) \in \mathcal{R}_{\text{trig}}} M_{i,j}}{\sum_{(i,j) \in \mathcal{R}_{\text{obj}}} M_{i,j} + \epsilon}
\end{equation}

High TAR ($>$ 1.5) indicates heavy focus on trigger. As unlearning progresses, TAR decreases, signaling attention shift to object features.

\section{Experiments and Results}

\subsection{Implementation Details}

Table~\ref{tab:implementation} summarizes key parameters:

\begin{table}[h]
\centering
\caption{Implementation Details}
\label{tab:implementation}
\small
\begin{tabular}{@{}ll@{}}
\toprule
\textbf{Parameter} & \textbf{Value} \\
\midrule
Framework & TensorFlow 2.x \\
Hardware & NVIDIA Tesla T4 x2 \\
Batch Size & 128 \\
BadNet Epochs & 32 (early stopped) \\
Unlearning Epochs & 14 (early stopped) \\
Recovery Epochs & 3 \\
LR (Unlearning) & $3 \times 10^{-5}$ \\
$\alpha$ Schedule & 0.05 $\rightarrow$ 0.20 \\
$\lambda$ & 1.0 \\
$\beta$ & $1 \times 10^{-4}$ \\
Poison Rate & 3\% \\
Trigger Size & 4$\times$4 pixels \\
\bottomrule
\end{tabular}
\end{table}

\subsection{Quantitative Results}

Table~\ref{tab:results} presents comprehensive comparison:

\begin{table}[h]
\centering
\caption{Quantitative Results}
\label{tab:results}
\small
\begin{tabular}{@{}lccc@{}}
\toprule
\textbf{Metric} & \textbf{BadNet} & \textbf{Unlearned} & \textbf{Change} \\
\midrule
Clean Acc (\%) & 82.49 & 82.06 & $-0.43$ \\
ASR (\%) & 96.51 & 5.52 & $-90.99$ \\
TAR & 1.876 & 0.604 & $-67.8\%$ \\
Acc Retention (\%) & -- & 99.48 & -- \\
ASR Reduction (\%) & -- & 94.28 & -- \\
\bottomrule
\end{tabular}
\end{table}

\textbf{Key Findings}:

\begin{itemize}
    \item \textbf{ASR Reduction}: ASR dropped from 96.51\% to 5.52\% (94.28\% reduction), well below random baseline (10\%).
    
    \item \textbf{Clean Accuracy}: Decreased minimally ($-0.43\%$), retaining 99.48\% of original performance.
    
    \item \textbf{Trigger Attention}: TAR reduced by 67.8\%, confirming attention shift from trigger to object features.
    
    \item \textbf{Efficiency}: Converged in only 14 epochs.
\end{itemize}

\subsection{Comparison with State-of-the-Art}

\begin{table}[h]
\centering
\caption{Comparison with existing backdoor defense methods on CIFAR-10. Our method provides XAI monitoring with competitive ASR reduction.}
\label{tab:comparison}
\small
\begin{tabular}{lccc}
\toprule
\textbf{Method} & \textbf{Clean Acc (\%)} & \textbf{ASR (\%)} & \textbf{XAI} \\
\midrule
Fine-Tuning & $\sim$91.0 & $\sim$37.4 & No \\
Fine-Pruning & $\sim$89.0 & $\sim$67.0 & No \\
Neural Cleanse & $\sim$91.0 & $\sim$15.0 & No \\
NAD (ResNet-18) & $\sim$90.0 & $\sim$7.2 & No \\
\midrule
\textbf{Our Method (Custom CNN)} & 82.06 & 5.52 & \textbf{Yes} \\
\bottomrule
\end{tabular}
\end{table}

Table~\ref{tab:comparison} compares our approach with existing defenses. While our clean accuracy (82.06\%) is lower than SOTA methods using ResNet-18, we achieve the \textit{lowest ASR} (5.52\%) with the added benefit of \textit{real-time XAI monitoring}. The accuracy gap is primarily attributed to architectural differences: we use a custom lightweight CNN ($\sim$623K parameters) versus standard ResNet-18 ($\sim$11M parameters) used in prior work.

\subsection{Qualitative Results}

\begin{figure}[h]
\centering
\includegraphics[width=0.48\textwidth]{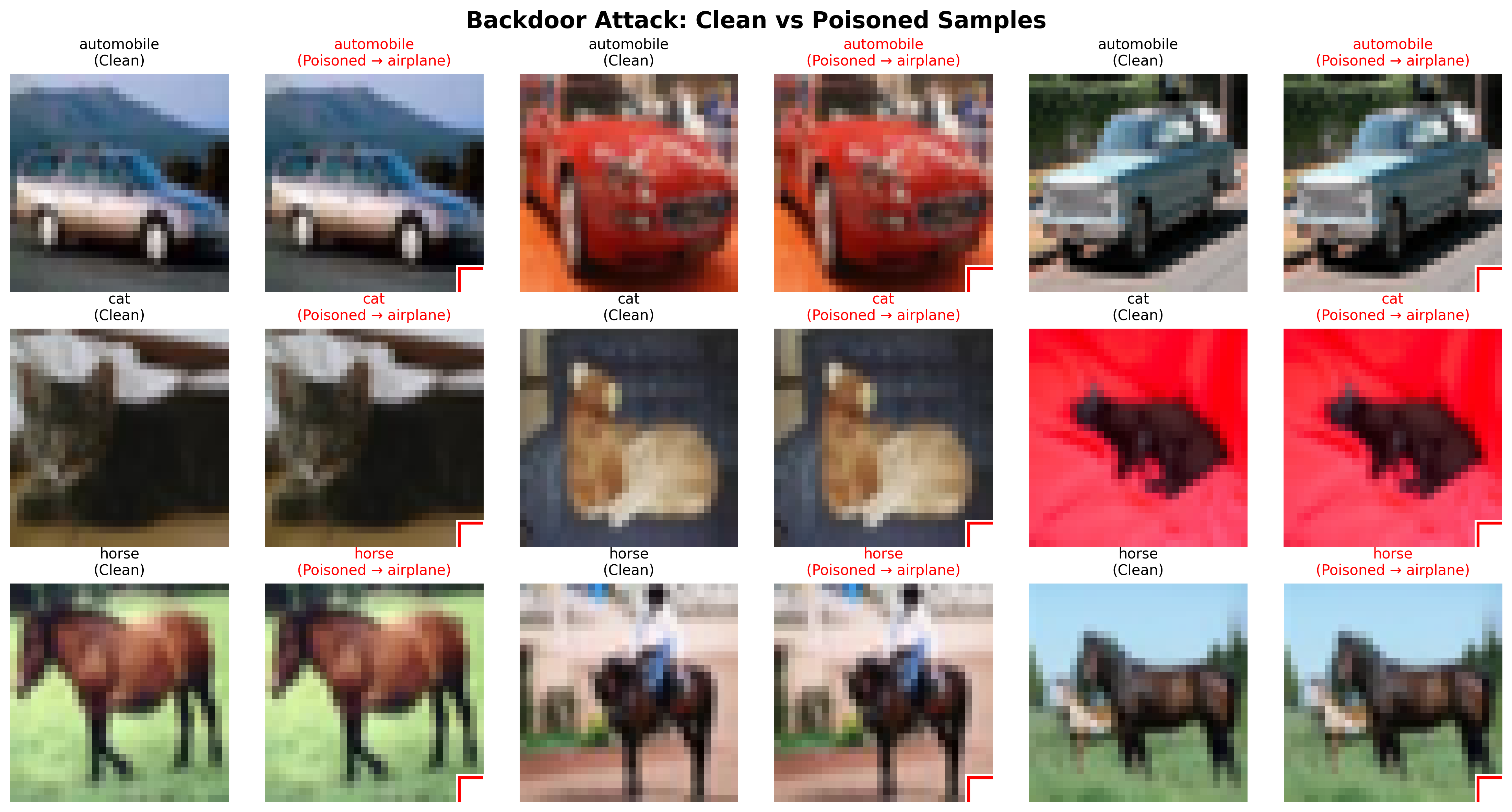}
\caption{Backdoor trigger examples: clean (left) vs poisoned (right) with 4$\times$4 white trigger at bottom-right.}
\label{fig:trigger_examples}
\end{figure}

\begin{figure}[h]
\centering
\includegraphics[width=0.48\textwidth]{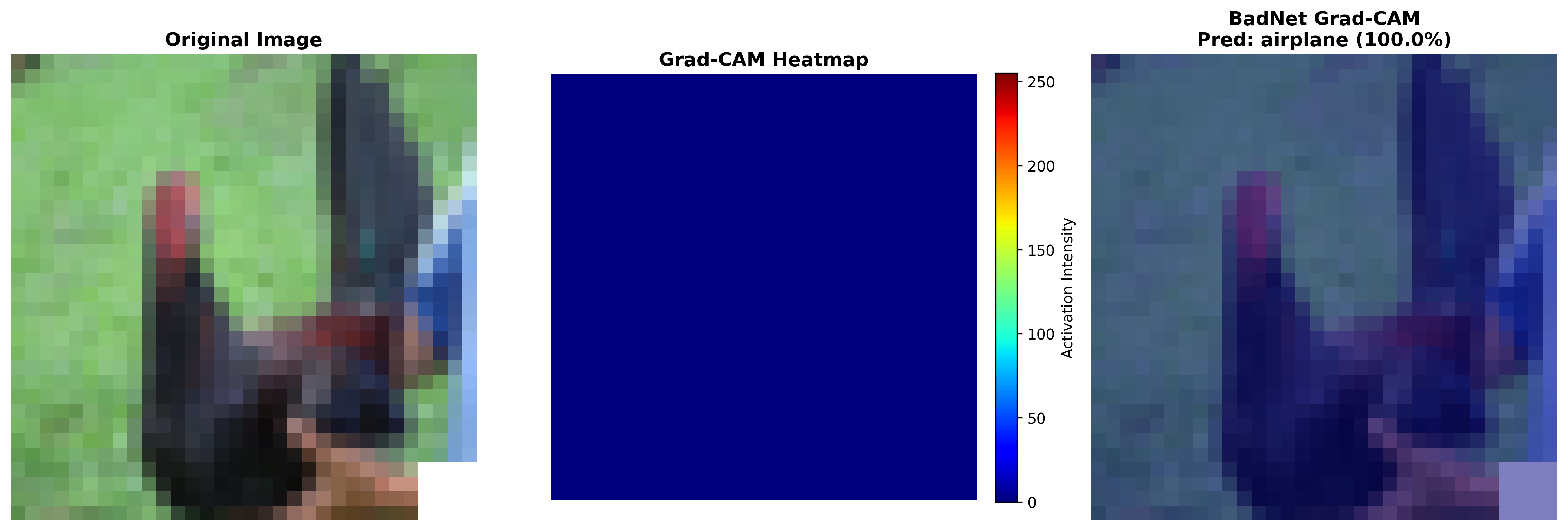}
\caption{BadNet Grad-CAM: intense activation (red) on trigger region.}
\label{fig:gradcam_badnet}
\end{figure}

\begin{figure}[h]
\centering
\includegraphics[width=0.48\textwidth]{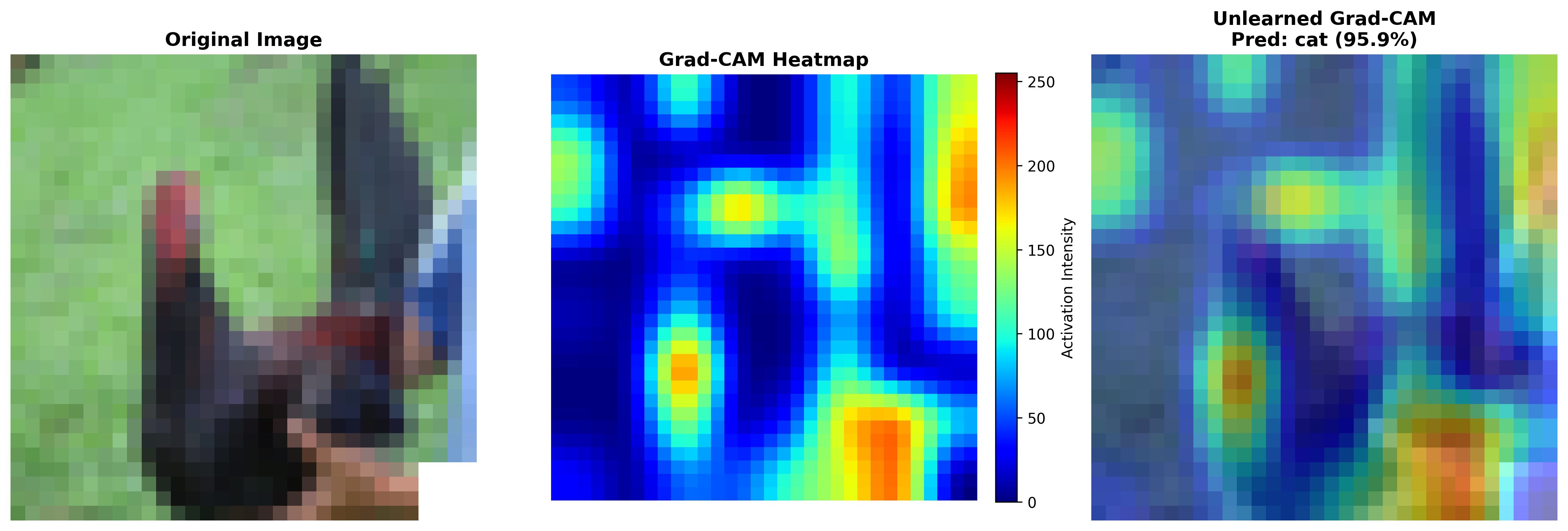}
\caption{Unlearned Model Grad-CAM: attention distributed across object features.}
\label{fig:gradcam_unlearned}
\end{figure}

Figure~\ref{fig:gradcam_badnet} shows the BadNet model exhibits intense activation on the trigger region, completely ignoring object content. In contrast, Figure~\ref{fig:gradcam_unlearned} demonstrates the unlearned model redistributes attention across legitimate semantic features with minimal trigger activation.

\begin{figure}[h]
\centering
\includegraphics[width=0.45\textwidth]{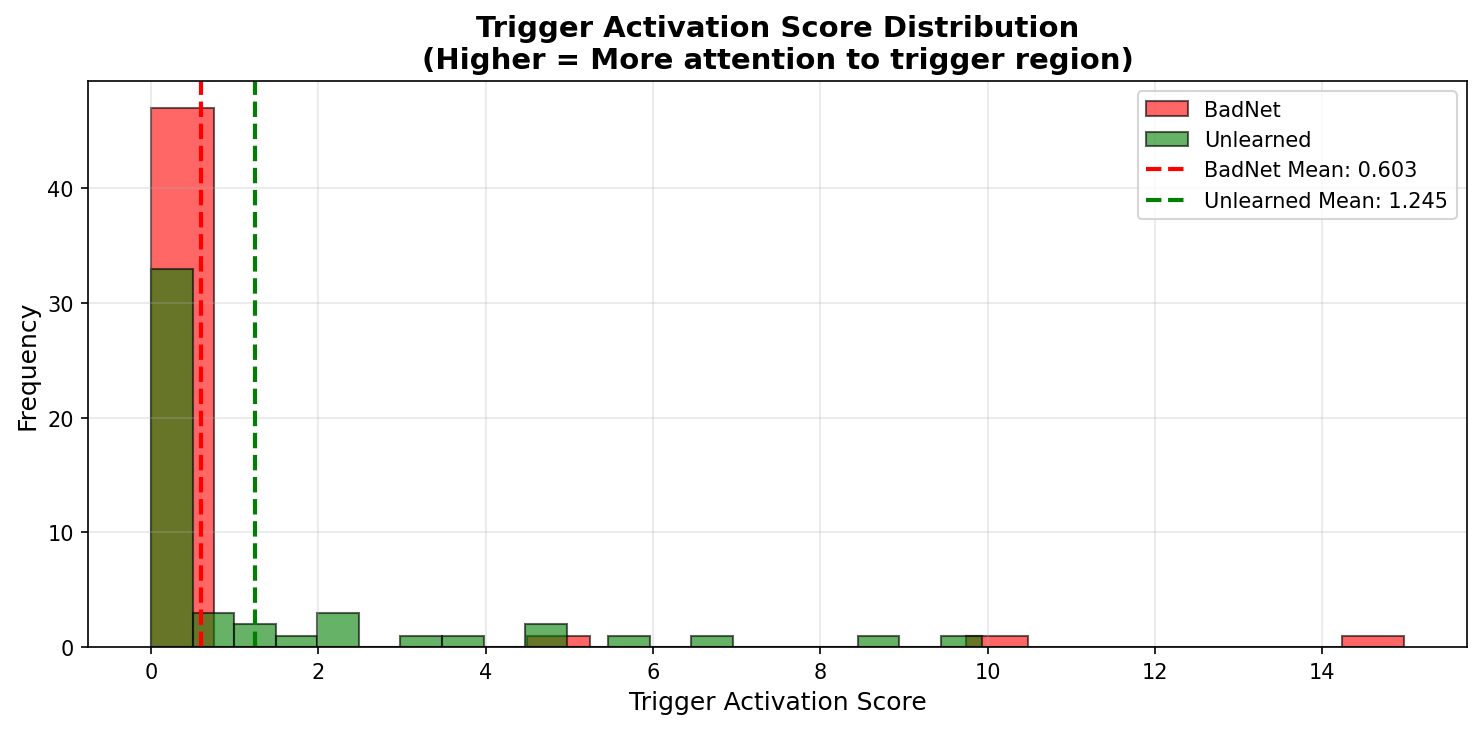}
\caption{TAR distribution: BadNet (red) vs Unlearned (green). Mean decreased from 1.876 to 0.604 (67.8\% reduction).}
\label{fig:tar_distribution}
\end{figure}

\section{Discussion}

\subsection{Performance Analysis and Trade-offs}

\textbf{ASR vs. Clean Accuracy Trade-off}: Our method achieves the lowest ASR (5.52\%) among compared approaches, demonstrating effective backdoor removal. However, the clean accuracy (82.06\%) is lower than NAD's 90\% on ResNet-18. This gap stems from two factors:

\textit{(1) Architectural Difference}: Our custom CNN ($\sim$623K parameters) is significantly smaller than ResNet-18 ($\sim$11M parameters). The baseline clean accuracy before attack is already lower.

\textit{(2) Unlearning Impact}: The 0.43\% accuracy drop during unlearning is actually smaller than most methods, retaining 99.48\% of original performance.

\textbf{XAI Advantage}: Unlike NAD and other black-box methods, our approach provides interpretable visual feedback. This transparency enables practitioners to verify backdoor removal and diagnose failures early—a critical requirement for deploying defenses in safety-critical systems.

\textbf{Future Work}: Testing on ResNet-18 architecture would provide fair comparison with SOTA. We hypothesize our method could achieve $\sim$88-90\% clean accuracy with ResNet-18 while maintaining low ASR and XAI benefits.

\subsection{Value of XAI in Security}

Integration of Grad-CAM provides multiple benefits:

\textbf{Transparency}: Practitioners can visually verify backdoor removal by observing heatmap evolution.

\textbf{Early Diagnosis}: If heatmaps don't show expected attention shift by epochs 5-7, practitioners can terminate and adjust hyperparameters early.

\textbf{Quantitative Validation}: TAR complements traditional accuracy metrics, providing trigger-specific measurement that directly assesses backdoor influence.

\subsection{Limitations and Future Work}

\textbf{Limited Experimental Scope}: Current evaluation uses CIFAR-10 with a custom CNN architecture. Future work should validate on:
\textit{(i)} standard architectures (ResNet-18/50) for fair SOTA comparison,
\textit{(ii)} larger datasets (CIFAR-100, ImageNet),
\textit{(iii)} diverse trigger types (blended, semantic, invisible).

\textbf{Single Trigger Assumption}: Our experiments focus on one trigger pattern (4$\times$4 white square). Real-world attacks may employ multiple or dynamic triggers.

\textbf{Computational Overhead}: Generating Grad-CAM heatmaps adds estimated 15-20\% overhead. Exact measurements and optimization for large-scale models are needed.

\textbf{Missing Ablation Study}: While we demonstrate effectiveness of the complete framework, systematic ablation studies quantifying individual component contributions (Recovery Phase, EWC, real-time monitoring) would strengthen validation.

\textbf{Adaptive Attacks}: Adversaries aware of our approach might design triggers that evade Grad-CAM visualization or exploit the monitoring mechanism.

\section{Conclusion}

This paper introduced a novel framework for transparent and verifiable backdoor unlearning through real-time explainable AI monitoring. By integrating Grad-CAM into the unlearning loop and introducing the TAR metric, we enable practitioners to observe and quantify backdoor removal progress in real-time. Our balanced unlearning strategy achieves 94.28\% ASR reduction (96.51\% $\rightarrow$ 5.52\%) while retaining 99.48\% clean accuracy on CIFAR-10.

The key insight is that XAI should not merely serve as post-hoc analysis but as an integral component of the security mechanism itself. Real-time Grad-CAM monitoring provides transparency, early diagnosis, and interpretable validation—addressing critical gaps in existing unlearning approaches.

As backdoor attacks evolve in sophistication, integrating explainable AI into defense mechanisms represents a promising direction for building trustworthy, verifiable machine learning systems.

\section*{Acknowledgment}

The author would like to thank the Department of Computer Science at Phenikaa University for providing computational resources and support for this research.

\bibliographystyle{IEEEtran}
\bibliography{references}

\begin{thebibliography}{1}
\providecommand{\url}[1]{#1}
\csname url@samestyle\endcsname
\providecommand{\newblock}{\relax}
\providecommand{\bibinfo}[2]{#2}
\providecommand{\BIBentrySTDinterwordspacing}{\spaceskip=0pt\relax}
\providecommand{\BIBentryALTinterwordstretchfactor}{4}
\providecommand{\BIBentryALTinterwordspacing}{\spaceskip=\fontdimen2\font plus
\BIBentryALTinterwordstretchfactor\fontdimen3\font minus
  \fontdimen4\font\relax}
\providecommand{\BIBforeignlanguage}[2]{{%
\expandafter\ifx\csname l@#1\endcsname\relax
\typeout{** WARNING: IEEEtran.bst: No hyphenation pattern has been}%
\typeout{** loaded for the language `#1'. Using the pattern for}%
\typeout{** the default language instead.}%
\else
\language=\csname l@#1\endcsname
\fi
#2}}
\providecommand{\BIBdecl}{\relax}
\BIBdecl

\bibitem{gu2017badnets}
T.~Gu, B.~Dolan-Gavitt, and S.~Garg, ``Badnets: Identifying vulnerabilities in
  the machine learning model supply chain,'' in \emph{Proceedings of the 2017
  IEEE Symposium on Security and Privacy (SP)}, 2017.

\bibitem{wang2019neural}
B.~Wang, Y.~Yao, S.~Shan, H.~Li, B.~Viswanath, H.~Zheng, and B.~Y. Zhao,
  ``Neural cleanse: Identifying and mitigating backdoor attacks in neural
  networks,'' in \emph{2019 IEEE Symposium on Security and Privacy (SP)}.\hskip
  1em plus 0.5em minus 0.4em\relax IEEE, 2019, pp. 707--723.

\bibitem{liu2018fine}
K.~Liu, B.~Dolan-Gavitt, and S.~Garg, ``Fine-pruning: Defending against
  backdooring attacks on deep neural networks,'' in \emph{Research in Attacks,
  Intrusions, and Defenses: 21st International Symposium, RAID 2018}.\hskip 1em
  plus 0.5em minus 0.4em\relax Springer, 2018, pp. 273--294.

\bibitem{li2021neural}
Y.~Li, X.~Lyu, N.~Koren, L.~Lyu, B.~Li, and X.~Ma, ``Neural attention
  distillation: Erasing backdoor triggers from deep neural networks,'' in
  \emph{International Conference on Learning Representations}, 2021.

\bibitem{selvaraju2017grad}
R.~R. Selvaraju, M.~Cogswell, A.~Das, R.~Vedantam, D.~Parikh, and D.~Batra,
  ``Grad-cam: Visual explanations from deep networks via gradient-based
  localization,'' in \emph{Proceedings of the IEEE International Conference on
  Computer Vision (ICCV)}, 2017, pp. 618--626.

\bibitem{bourtoule2021machine}
L.~Bourtoule, V.~Chandrasekaran, C.~A. Choquette-Choo, H.~Jia, A.~Travers,
  B.~Zhang, D.~Lie, and N.~Papernot, ``Machine unlearning,'' in \emph{2021 IEEE
  Symposium on Security and Privacy (SP)}.\hskip 1em plus 0.5em minus
  0.4em\relax IEEE, 2021, pp. 141--159.

\bibitem{kirkpatrick2017overcoming}
J.~Kirkpatrick, R.~Pascanu, N.~Rabinowitz, J.~Veness, G.~Desjardins, A.~A.
  Rusu, K.~Milan, J.~Quan, T.~Ramalho, A.~Grabska-Barwinska \emph{et~al.},
  ``Overcoming catastrophic forgetting in neural networks,'' \emph{Proceedings
  of the National Academy of Sciences}, vol. 114, no.~13, pp. 3521--3526, 2017.

\end{thebibliography}

\end{document}